\documentclass[twocolumn,secnumarabic,amssymb, nobibnotes, aps]{revtex4-1}

\usepackage{graphicx}
\usepackage{subfigure}
\usepackage{float}
\usepackage{color}

\begin{document}

\title{Investigation of electronic trap states in organic  photovoltaic materials by current-based deep level transient spectroscopy}

\author{Stefan Neugebauer $^1$}
\author{Julia Rauh $^1$}
\email[Electronic mail: ]{julia.rauh@physik.uni-wuerzburg.de}
\author{Carsten Deibel $^1$}
\author{Vladimir Dyakonov $^{1,2}$}
\email[Electronic mail: ]{dyakonov@physik.uni-wuerzburg.de}
\affiliation{$^1$ Experimental Physics VI, Faculty of Physics and Astronomy, Julius-Maximilians University of W\"urzburg, Am Hubland, 97074 W\"urzburg, Germany}
\affiliation{$^2$ Bavarian Center for Applied Energy Research (ZAE Bayern), Am Hubland, 97074 W\"urzburg, Germany}

\begin{abstract}
Current-based deep level transient spectroscopy was used to study trap states in poly(3-hexylthiophene-2,5-diyl) (P3HT), [6,6]-phenyl-C61 butyric acid methyl ester (PCBM) and P3HT:PCBM blend. The obtained spectra showed traps of 87~meV activation energy in pure P3HT and 21~meV for PCBM. The blend shows a complex emission rate spectrum consisting of several different emission rate bands in the range of (0.1--30)~s$^{-1}$, yielding activation energies between about 30~meV and 160~meV. 

\end{abstract}

\maketitle

\clearpage

Recombination of charge carriers is one important process limiting the performance of organic solar cells, which have reached efficiencies in power conversion of up to 8.3~\%~\cite{Green2011}. However, it is still a controversial question about the dominant recombination process which restricts photocurrent. This controversy also includes the possible role of trap states among these loss processes.
Trapped charge carriers are immobile, representing a recombination centre for free charge carriers but can be released by thermal activation. So two processes involving trapped charge carriers are possible: Recombination of a free and a trapped charge carrier which is a first order recombination process, whereas delayed bimolecular recombination of thermally emitted trapped charge carriers may be considered as a recombination process of the second and even higher order~\cite{Deibel2010}.
For modeling recombination occurring in the solar cells, data about the energetic distribution and -- especially in case of the bimolecular recombination process -- about the emission rates of trap states are important.

In this work we used current-based deep level transient spectroscopy (I-DLTS) to examine trap states in P3HT (poly(3-hexylthiophene-2,5-diyl)), PCBM ([6,6]-phenyl-C61 butyric acid methyl ester) and P3HT:PCBM blends. 

The P3HT, PCBM and P3HT:PCBM (weight ratio 1:0.8) samples were fabricated in nitrogen atmosphere using spin coating procedure on poly(3,4-ethylenedioxythiophene): (polystyrenesulfonate) (PEDOT:PSS) covered ITO glass substrates. Pure P3HT and P3HT:PCBM blends were processed from 30~mg/ml chlorobenzene solution, in case of pure PCBM chloroform solution (20~mg/ml) was used.
The resulting film thicknesses were about 190~nm for P3HT, 140~nm for PCBM and 150~nm for P3HT:PCBM blends, as measured by a profilometer.
The blends were annealed for 10~min at 130~$^\circ$C. No thermal treatment was applied to pure P3HT and PCBM.
For P3HT and P3HT:PCBM blends Ca/Al (3~nm/120~nm) top electrodes were used and LiF/Al (1~nm/120~nm) for PCBM, which were thermally evaporated at a base pressure of $< 10^{-6}$~mbar.
The effective sample areas were about 3~mm$^2$. All materials were used without further purification. P3HT was purchased from Rieke Metals (Sepiolid~P200), PCBM (purity $\ge$ 99.5~\%) from Solenne and PEDOT:PSS from H.~C.~Starck (CLEVIOS~P~VP~AI~4083).

I-DLTS measurements were performed in a closed cycle cryostat with helium as contact gas. 
Via an integrated lock, the samples were transferred from the glovebox to the cryostat, which avoided potential degradation of the samples due to air exposure.
For trap filling a voltage pulse of 3~V for 6~s in forward direction was used. These conditions were found to be optimal in terms of saturation of the I-DLTS signal. 
 We point out that varying the pulse heights (0.5~V--3.5~V) and widths (0.2~s--10~s) resulted only in a change of the amplitude of the I-DLTS signal and had almost no effect on the distribution or width of the obtained emission rate spectra.
The current transients were recorded for 80~s, as long as  the current resolution limit of $10^{-12}$~A was reached. We note, five orders of magnitude of current variation and four orders of magnitude in time were experimentally accessible. The measurements were performed in the temperature range between 62~K and 306~K in steps of 2~K.
During the transients no external electric field was applied to avoid injection current, implying that the de-trapped charge carriers were extracted from the samples only due to the built-in voltage. 
The injection voltage pulse was applied to the samples using an ADwin-Gold D/A-converter.  Transient current signals were amplified by a FEMTO-DLPCA-200 and recorded by the ADwin-Gold. 

 The basic principle of the I-DLTS measurement can be briefly summarized as follows: The sample, being in thermal equilibrium at the beginning of the measurement, is perturbed by a positive voltage pulse injecting charge carriers. This leads to an occupation of trap states. Switching the external voltage off, the sample relaxes back to steady-state condition by emission of the trapped charge carriers, resulting in a current transient. Thus, the decay of the current transient is related to the emission rates of the trapped charges.
Since emission of charge carriers out of trap states is assumed to be thermally activated with activation energy $\Delta E$, the emission rate $e_T$ at temperature $T$ from a discrete energy level is given by:  
\begin{equation}
	\centering
		e_T=\nu_0\exp\left(-\frac{\Delta E}{k_BT}\right)~~~,
	\label{eq:eact}
\end{equation}
where $k_B$ is the Boltzmann constant and $\nu_0$ a prefactor.

Thus, from the Arrhenius plot of the emission rates the activation energies can be extracted. Thereby, in the case of organic semiconductors, either $e_T/T^2$ \cite{Gaudin2001,Nguyen2006a} or $e_T$~\cite{Jones1997a} is plotted over $1/T$. The former is in analogy to the analysis of inorganic semiconductors,
which show temperature dependencies of the effective density of states $\propto T^{3/2}$ and the thermal velocity $\propto T^{1/2}$ resulting in an additional $T^2$ dependence of the prefactor. In the latter, the temperature dependence of the prefactor is completely neglected, in analogy to the Miller--Abrahams hopping rate equation~\cite{miller1960}, often used to describe charge transport in organic semiconductors~\cite{Grunewald1979,schmechel2003,arkhipov2001a}.
According to this simplification, we also neglected a temperature dependence of the prefactor here.

In the case of a distribution $N_T(E)$ of trap states and consequently a distribution $N_T(e_T)$ of emission rates $e_T$, the current transient $I(t)$ is given by: 
\begin{equation}
I(t)\propto \int_0^\infty N_T(e_T) e_T  \exp(-t e_T) de_T~~~.
\label{eq:cur}	
\end{equation}
Further details about DLTS theory can be found in literature~\cite{lang1974,arora1993,stallinga2000}. 

 \begin{figure}	
	\includegraphics{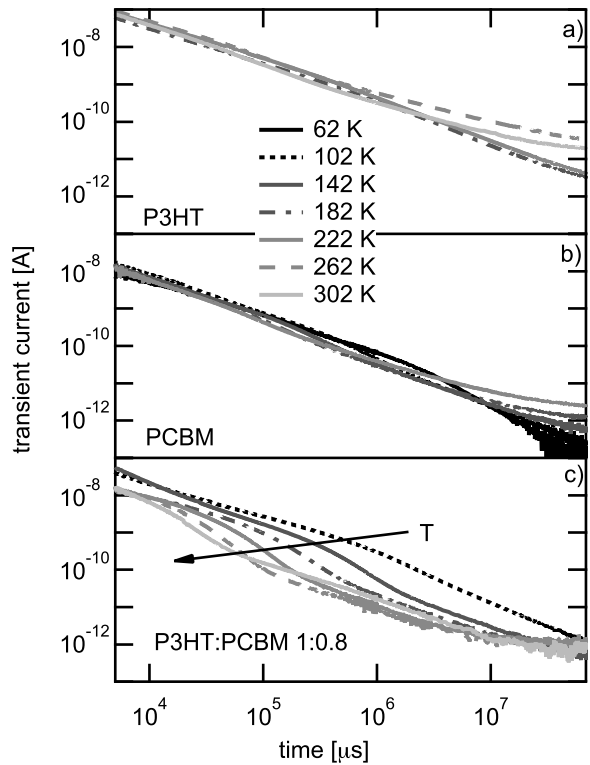}
	\caption{Exemplary current transients of P3HT, PCBM and P3HT:PCBM blend  after a trap filling voltage pulse of 3~V for 6~s. During transient recording no external voltage was applied.}
	\label{fig:Fig1}
\end{figure}

In Fig.~\ref{fig:Fig1} exemplary transient currents for P3HT, PCBM and P3HT:PCBM are shown for those temperatures where the thermally dependent emission rates were obtained.
The transients of the pure materials have slower declines for the higher temperatures than those of the blend but have weaker temperature dependence. The blend however shows the most complex transients with decay times of strong temperature dependence indicating several thermal emitting states with high activation energies.

The emission rate spectra $N_T(e_T)$ were calculated from the current transients using the program FTIKREG~\cite{Weese1993,Weese1992}, which is an implementation of the Tikhonov regularization method, providing a much better resolution than the common boxcar method often used in literature~\cite{Gaudin2001,lang1974,thurzo2003,Nguyen2004,Nguyen2006a,Huang2007,renaud2008a}. Moreover boxcar was initially developed to find decay times of mono-exponential transients, while FTIKREG can be used to analyze transients which contain distributions of emission rates~\cite{Istratov1999a}. 

\begin{figure}	
	\includegraphics{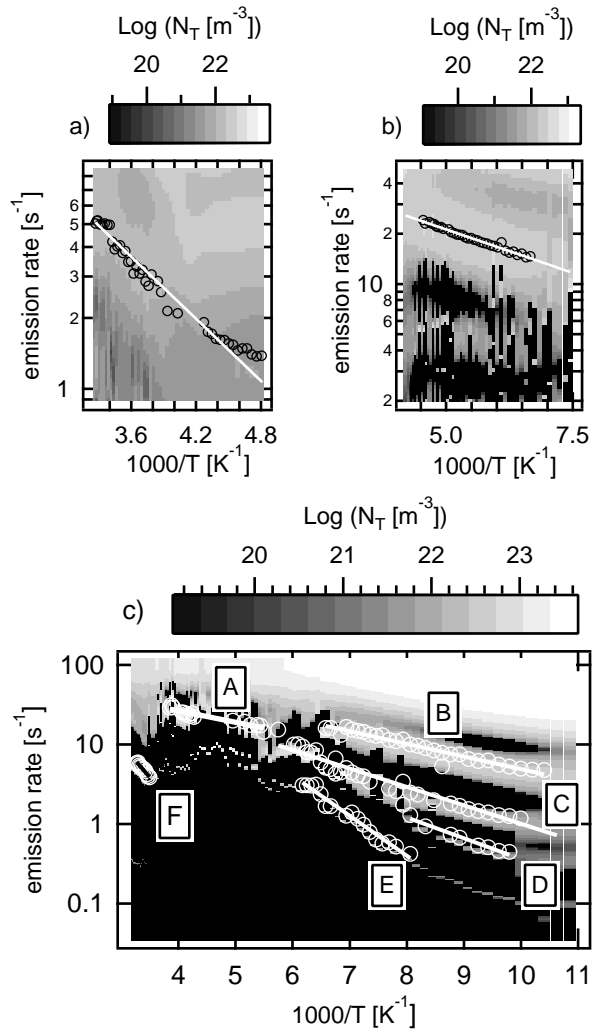}
	\caption{Emission rate spectra of pure P3HT a), PCBM b) and P3HT:PCBM blend c). The exact maxima of emission rate distributions were determined by multiple peak fit (circles). In addition the Arrhenius-fits to identify the activation energies are shown (lines).}
	\label{fig:Fig2}
\end{figure}	
	
The obtained emission rate spectra of P3HT, PCBM and the P3HT:PCBM blend for the different temperatures are shown in Fig.~\ref{fig:Fig2} as Arrhenius plots. The colors of the contour plot represent the amplitudes of the emission rate distributions with $N_T$ refering to the density of extracted charge carriers per emission rate window, where the boundaries of the rate windows are equidistant on the logarithmic axis. According to equation~\ref{eq:eact}, the activation energies of the corresponding trap states can be extracted from Fig.~\ref{fig:Fig2} by linear fits of maxima (circles) of the emission rate spectra. To determine the exact positions of the maxima, the emission rate spectra were fitted separately for each temperature by applying multiple peak fits with Gaussian shapes.

The spectra of pure P3HT and PCBM show more charge carriers emitted at lower rates and a weaker temperature dependence compared to the blend (Fig.~\ref{fig:Fig2}), as it was expected in consequence of the slower decays for the transients of the pure materials. 

As can be seen in Fig.~\ref{fig:Fig2}~a) the emission rate spectra of P3HT contain an emission rate band in the range of (1--5)~s$^{-1}$ at high temperatures between 230 and 306~K, whereas at lower temperatures no temperature dependence of the emission rates can be observed. The  corresponding activation energy of the trap is  87~meV, as obtained by the Arrhenius fit (line).

Fig.~\ref{fig:Fig2}~b) shows an emission rate band of PCBM in the range of (10--30)~s$^{-1}$ at temperatures between 180 and 220~K indicating shallow states with an activation energy of 21~meV.

\begin{figure}
	  \includegraphics{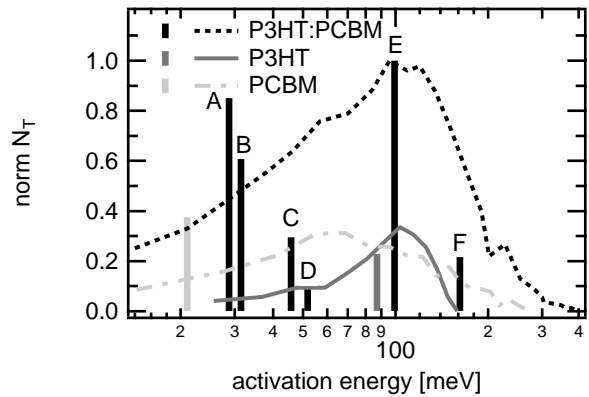}
	\caption{Bars: Overview of the obtained activation energies in PCBM (light grey), P3HT (dark grey) and P3HT:PCBM blend (black) and relative maximum $N_T$ in the spectrum found by I-DLTS. Lines: normalized activation energy spectra of earlier TSC measurements~\cite{Schafferhans2010,schafferhans2008,Schafferhans2011} (colors are as above).}
	\label{fig:Fig3}	
\end{figure}	

The emission rate spectra of P3HT:PCBM blend, shown in Fig.~\ref{fig:Fig2}~c), consist of several different emission rate bands in the range of (0.1--30)~s$^{-1}$, which are separated by areas of zero amplitude. 
Thereby, the emission rate band marked with A is in good agreement with that obtained for the pure PCBM, yielding also a similar activation energy of 28~meV. The emission rate band of pure P3HT, however, is not detected in the blend. This might be due to the overlap with the more pronounced emission rate distribution marked as F in the blend, and thus, a limited resolution. The emission rate band F is strongly temperature dependent, resulting in an activation energy of the corresponding traps of about 160~meV. 
Furthermore, the blend features additional emission rate bands marked with B--E exhibiting similar temperature dependencies as in pure PCBM and P3HT. Therefore, they yield similar activation energies between 30~meV and 100~meV. However, these traps are not observed in pure P3HT or PCBM, they are only found in the blend, which exhibits a higher energetic disorder due to the intermixing of both materials. In Fig.~\ref{fig:Fig3} the activation energies and the relative amplitudes (i.e. the emission rate amplitudes normalized to the maximum amplitude) of the three different samples are summarized.

Furthermore Fig.~\ref{fig:Fig3} shows results of earlier TSC (thermally stimulated current) studies~\cite{schafferhans2008,Schafferhans2010,Schafferhans2011} for comparison. 
All obtained activation energies by I-DLTS are in good accordance with TSC. The emission rate band E of the blend and the activation energy of P3HT fits the peak positions of the TSC spectra very well. Both measurements show additional deeper trap states that exclusively exist in the blend and significantly higher densities of trapped charge carriers in the blend than in the pure materials.

Peaks in emission rate spectra of DLTS measurements of organic materials~\cite{renaud2008a} are often fitted by several Gaussian peaks due to the low resolution of the common boxcar method. 
Since DLTS gives the opportunity to analyze emission rates independently of the corresponding activation energies, trap states of similar energetic depth but different emission rates can be clearly  
distinguished if the transients are analyzed applying the Tikhonov regularization providing sufficient resolution in emission rates. 
This is important as small differences in activation energies result in large variations of emission rates, especially at low temperatures, in consequence of the exponential dependency. Moreover since there is no absolute value given for the prefactor in equation \ref{eq:eact} it is not possible to calculate emission rates for a definite activation energy.
The emission rate bands B, C and D have similar slopes, appear in similar temperature regimes and are covered by the TSC spectrum of the blend, but their emission rates differ more than one order of magnitude, as can be seen in Fig.~\ref{fig:Fig2}~c). 
Information on emission rates of trap states is of fundamental importance to describe charge carrier dynamics in case of transient experiments. Trapped charge carriers act -- after they have been thermally emitted -- as free charge carriers and thus can participate in bimolecular recombination. Current studies of recombination showed bimolecular recombination with order larger than two~\cite{shuttle2008,Shuttle2008a,Juska2008}. This delayed recombination of charge carriers can be attributed to sub-band-gap localized trap states in the exponential tail of the density of states~\cite{Foertig2009,shuttle2008} or alternatively to extrinsic trap states.

In conclusion we applied current based I-DLTS technique to P3HT:PCBM blend solar cells as well as to P3HT and PCBM using Tikhonov regularization to achieve emission rate spectra in high resolution. We found several emission rate bands with trap activation energies of 87~meV in pure P3HT, 21~meV for PCBM and a trap level distribution in the blend between about 30~meV and 160~meV. Moreover the blend shows complex emission rate spectra with bands at different emission rates but similar activation energies. This shows the importance of further studies of trap states focusing on emission rate spectra to improve understanding and modeling of bimolecular charge carrier recombination. 

The current work is supported by the Bundesministerium f\"ur Bildung und Forschung in the framework of the OPV Stability Project 
(Contract No. 03SF0334F). C.D. gratefully acknowledges the support of the Bavarian Academy of Sciences and Humanities. V.D.'s work at the ZAE Bayern is financed by the Bavarian Ministry of Economic Affairs, Infrastructure, Transport and Technology.

\end{document}